\documentclass[%
reprint,
superscriptaddress,
nofootinbib,
longbibliography,
amsmath,
amssymb,
aps,
showkeys
]{revtex4-2}
\usepackage{graphicx}
\usepackage{dcolumn}
\usepackage{bm}
\usepackage{hyperref}
\usepackage[bottom]{footmisc}
\usepackage{footnote}
\usepackage{xcolor}
\usepackage{booktabs}
\usepackage{array,multirow}
\usepackage{amsmath}
\usepackage{amsfonts}
\usepackage{amssymb}
\usepackage{mathrsfs}
\usepackage{enumerate}
\usepackage{float}
\usepackage{verbatim}
\usepackage{braket}
\usepackage{setspace}
\usepackage[normalem]{ulem}
\usepackage{multirow}
\usepackage{comment}
\usepackage{tikz}
\usepackage{bm}
\usepackage{dsfont}
\usetikzlibrary{calc}
\allowdisplaybreaks
\hypersetup{breaklinks=true, colorlinks=true, citecolor=blue, linkcolor=blue, urlcolor=blue}

\begin{document}

\newcommand{\LAL}{$\Lambda\bar{\Lambda}$}
\newcommand{\PLal}{P_{\Lambda\bar{\Lambda}}}
\newcommand{\DR}{\Delta R}
\newcommand{\Dy}{\Delta y}
\newcommand{\Dphi}{\Delta\phi}
\newcommand{\sqs}{\sqrt{s}}
\newcommand{\dNdy}{\mathrm{d}N/\mathrm{d}y}
\newcommand{\Npair}{N_\mathrm{pair}}
\newcommand{\chisq}{\chi^2}
\newcommand{\ffd}{f_\mathrm{d}}

\setlength{\parskip}{0pt}
\setlength{\parindent}{1em}

\newcommand{\resultbox}[2]{%
  \medskip\noindent\textbf{#1.}\ \itshape #2\upshape\medskip}

\title{Qubit--Qutrit Quantum Tomography of hadronic $\Lambda\phi$ and $\Lambda K^{\ast 0}$ systems}

\author{Veronica~Kanachova}
\affiliation{Department of Physics and Astronomy, Haverford College,
             Haverford, Pennsylvania 19041, USA}

\author{Feng Liu}
\affiliation{Department of Physics and Astronomy, Stony Brook University, Stony Brook,
             New York 11794-3800, USA}

\author{Zhoudunming~Tu}
\email{zhoudunming@bnl.gov}
\affiliation{Department of Physics and Astronomy, Stony Brook University, Stony Brook,
             New York 11794-3800, USA}
\affiliation{Department of Physics, Brookhaven National Laboratory,
             Upton, New York 11973-5000, USA}

\date{\today}

\begin{abstract}
Quantum-information observables have emerged in recent years as new tools in nuclear and particle physics, from entanglement in top-quark pairs to spin correlations in $\Lambda\bar{\Lambda}$ production. Extending these studies to unequal-spin hadronic final states poses a fundamental challenge: the $6\times6$ density matrix of a qubit--qutrit system contains 35 independent spin parameters, but the decays of $\Lambda V$ pairs, with $V=\phi$ or $K^{*0}$, provide access to only 23 due to the hidden vector polarization from the strong decay. In this Letter, we formulate a qubit--qutrit quantum tomography (QQQT) technique for these spin-$\tfrac{1}{2}\otimes1$ systems and establish exact criteria for entanglement certification from the \textit{incomplete} density matrix. Compared with the $\Lambda\bar{\Lambda}$ system, QQQT of $\Lambda\phi$ and $\Lambda K^{*0}$ provides a new probe of nonperturbative QCD hadronization, enabling a direct comparison of the spin evolution of entangled quark pairs produced from the vacuum as they hadronize into a baryon or a vector meson.
\end{abstract}

\maketitle

\textbf{\textit{Introduction.}} How quarks and gluons form hadrons remains one of the least understood aspects of Quantum Chromodynamics (QCD). Known as hadronization, it has been modeled phenomenologically, through string or cluster fragmentation tuned to data, rather than derived directly from the first principles of confining dynamics. An especially open question concerns the role of \textit{spin}: whether, and how, spin correlations generated among partons at short distances survive, evolve, or decohere as the system enters the nonperturbative regime and forms color-singlet hadrons.

Quantum-information observables, including entanglement, Bell nonlocality, and quantum-state tomography, have recently emerged as quantitative probes of such spin correlations~\cite{EPR1935,Bell1964,CHSH1969,BarrEtAl2024,AfikEPPSU2025,KharzeevQI2021}. The ATLAS Collaboration has observed spin entanglement in top-quark pairs at the LHC~\cite{ATLAS2024}, while complementary studies have reconstructed the top-quark spin density matrix and explored quantum-state tomography~\cite{AfikMunoz2021,CMS2019SpinDensity,CMS2024SpinTomography,CMSljets2024,AfikMunozQuantum2022,ChengHanLow2024}. Because top quarks decay before hadronization, these correlations are established at the partonic level. By contrast, the STAR Collaboration has reported an analogous spin-correlation measurement for $\Lambda\bar{\Lambda}$ hyperon pairs produced in proton-proton collisions~\cite{STARNature2026}, extending such quantum observables to hadronic final states~\cite{Tornqvist1981,GongParidaTuVenugopalan2022,ShiYang2019,BESIII2019,BESIII2025LocalRealism,ThrustCut2025,AfikBottom2025}. In this case, the correlations must persist through the nonperturbative formation of two baryons. Taken together, these measurements probe spin correlations on opposite sides of hadronization: before confinement can act on the spin state and after hadron formation is complete.

The systems studied so far, however, involve particles of equal spin: both $t\bar t$ and $\Lambda\bar{\Lambda}$ are spin-$\tfrac12\otimes\tfrac12$ systems. A natural next step is therefore a mixed-spin system, consisting of a spin-$\tfrac12$ hyperon produced together with a spin-1 vector meson ($V$), such as $\Lambda\phi$ or $\Lambda K^{*0}$. There is also a physics motivation - such final states provide a qualitatively new probe of spin transport through hadronization. A potentially correlated quark pair excited from the QCD vacuum with maximal entanglement~\cite{STARNature2026,LiuTu2026,HyperonHI2025} (or any other scenarios, e.g., jet fragmentation) can evolve through two structurally distinct hadron-formation channels, producing a baryon and a vector meson~\cite{LiuTu2026,BarataEtAl2026,AnderssonLund1983,Webber1984,Pythia83,vonKuk2025,ZhangWei2023,ChenLiangSongWei2022}. Comparing these two channels offers a new way to investigate how quark-spin correlations are propagated into hadronic degrees of freedom during hadronization.

\textbf{\textit{Challenge.}} The central experimental challenge is incomplete tomography. In principle, the relevant tools from Quantum Information Science (QIS) can be applied directly: for a qubit--qutrit state in a $2\times3$ Hilbert space, the positive-partial-transpose (PPT) criterion is both necessary and sufficient for separability. Experimentally, however, not all components of the density matrix are accessible. The weak decay of the $\Lambda$ is self-analyzing, whereas the strong decay of the vector meson is insensitive to part of its polarization. Consequently, only 23 of the 35 independent spin parameters of the full qubit--qutrit density matrix can be measured, leaving 12 experimentally inaccessible.

In this Letter, we formulate a qubit--qutrit quantum tomography (QQQT) technique for $\Lambda V$ systems, construct the most general spin density matrix, and determine precisely which of its components are experimentally accessible. We then establish exact conditions under which entanglement can be inferred from the incomplete density matrix and quantify the limitations imposed by the unmeasured sector. These results provide both the theoretical foundation and the experimental framework for entanglement measures in hadronic qubit--qutrit systems, which establish $\Lambda\phi$ and $\Lambda K^{*0}$ production as new probes of spin evolution in nonperturbative QCD.

\begin{figure}[t]
    \centering
    \includegraphics[width=0.49\textwidth]{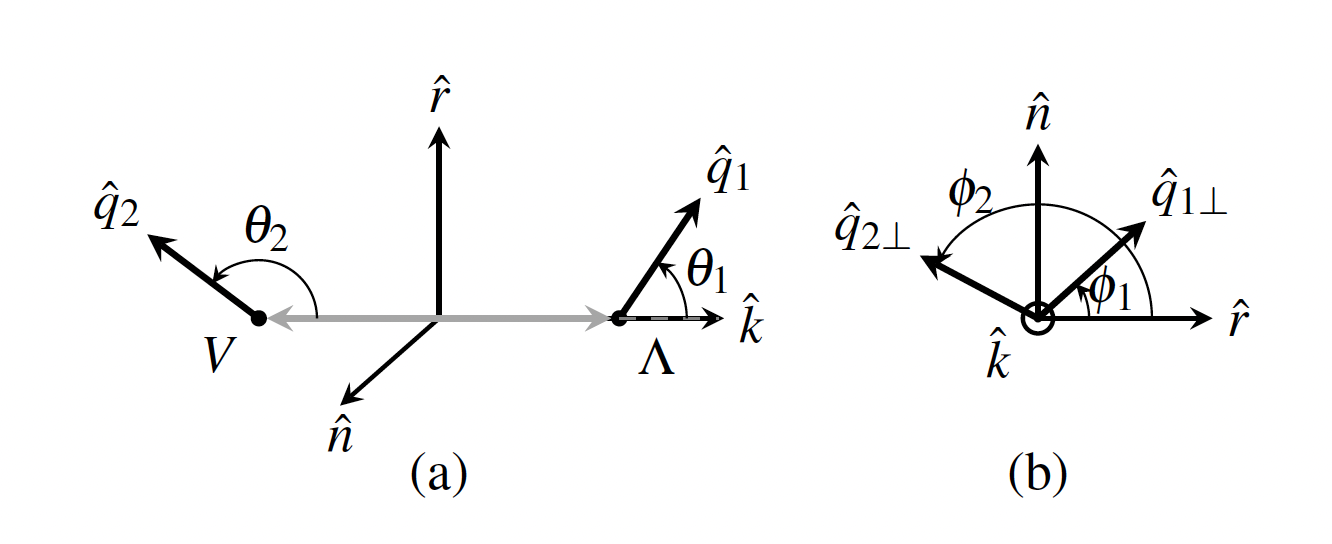}
\caption{Coordinate system, defined in the $\Lambda V$ pair rest frame, where the two hadrons are back to back. The orthonormal triad is $\hat k=\hat p_{\,\Lambda}$ (the $\Lambda$ momentum direction), $\hat n\propto\hat p_{\rm beam}\times\hat k$ normal to the production plane, and $\hat r=\hat n\times\hat k$, with $\hat k\times\hat r=\hat n$. The daughter directions $\hat q_1$ (proton from $\Lambda\to p\pi^-$) and $\hat q_2$ ($K^+$ from $\phi\to K^+K^-$ or $K^{*0}\to K^+\pi^-$) are each evaluated in the rest frame of their own parent but referred to this single shared triad.
}
\label{fig:helicity}
\end{figure}

\textbf{\textit{Formalism.}} We define the spin quantization axes in the $\Lambda V$ pair rest frame using the orthonormal triad $(\hat k,\hat r,\hat n)$ shown in Fig.~\ref{fig:helicity}, with $\hat k$ along the $\Lambda$ momentum, $\hat n$ normal to the production plane, and $\hat r=\hat n\times\hat k$. Throughout, we identify $(\hat x,\hat y,\hat z)\equiv(\hat r,\hat n,\hat k)$. This is the standard helicity basis used in $t\bar t$ analyses~\cite{CMS2024SpinTomography,AfikMunoz2021,BarrEtAl2024}. Since the two hadrons are back-to-back in the pair rest frame, a single triad suffices to describe both spin states.

We first consider the two particles separately. The density matrix of the spin-$\tfrac12$ $\Lambda$ is
$\rho_A=\tfrac12(\mathds{1}_2+a_i\sigma_i)$,
where $a_i=\langle\sigma_i\rangle$ are the three components of the $\Lambda$ polarization. A spin-1 vector meson requires, in addition to its vector polarization, a rank-2 tensor describing its spin alignment. Introducing the spin-1 generators $S_i$ in the helicity basis ${|{+1}\rangle,|0\rangle,|{-1}\rangle}$ and the traceless quadrupole tensor
$T_{jk}\equiv\tfrac32(S_jS_k+S_kS_j)-2\delta_{jk}\mathds{1}_3$,
the vector-meson density matrix can be written as,
\begin{align}
\rho_B={}&\tfrac13\mathds{1}_3+\tfrac12P_iS_i+\big[\rho_{00}T_{zz}+R_{10}T_{xz}\notag\\
&+R_{1m}T_{xx-yy}+I_{10}T_{yz}+I_{1m}T_{xy}\big],
\end{align}
where $P_i=\langle S_i\rangle$ denotes the vector (dipole) polarization. The five parameters in brackets describe the rank-2 quadrupole alignment familiar from spin-alignment measurements of $\phi$, $K^{*0}$, and $J/\psi$~\cite{SchillingWolf1970,STARSpinAlignment,ShengPhiField2023,XuHuang2024,ShengReview2024}. Up to normalization conventions, they correspond to the helicity-density-matrix elements $\rho_{00}$, $\mathrm{Re}(\rho_{1,0}$), $\mathrm{Re}(\rho_{1,-1}$), $\mathrm{Im}(\rho_{1,0}$), and $\mathrm{Im}(\rho_{1,-1}$) of the Schilling--Wolf parametrization. Together, the three dipole and five quadrupole components provide the eight independent parameters of a spin-1 density matrix.

We now combine the two spin states. In the operator basis
${\mathds{1}_2,\sigma_i}\otimes{\mathds{1}_3,S_i,T_{jk}}$,
the joint $2\times3=6$ dimensional Hilbert space carries a Hermitian, unit-trace density matrix with $6^2-1=35$ independent real parameters,
\begin{align}
\rho={}&\tfrac16\Big[\mathds{1}_2\!\otimes\!\mathds{1}_3+a_i(\sigma_i\!\otimes\!\mathds{1}_3)+\mathds{1}_2\!\otimes\!\big(\rho_{00}T_{zz}\notag\\
&+R_{10}T_{xz}+R_{1m}T_{xx-yy}+I_{10}T_{yz}+I_{1m}T_{xy}\big)\notag\\
&+\mathds{1}_2\!\otimes\!P_iS_i+C_{ij}(\sigma_i\!\otimes\!S_j)\notag\\
&+K_{i,\alpha}(\sigma_i\!\otimes\!T_{\alpha^{\rm even}})+L_{i,\alpha}(\sigma_i\!\otimes\!T_{\alpha^{\rm odd}})\Big],
\label{eq:rhofull}
\end{align}
with $\alpha^{\rm even}\!\in\!\{zz,xz,xx\!-\!yy\}$ and $\alpha^{\rm odd}\!\in\!\{yz,xy\}$. The 35 parameters have a natural physical organization: three describe the $\Lambda$ polarization, eight describe the vector-meson polarization and alignment, and the remaining 24 encode correlations between the two spins. The latter consist of nine dipole--dipole coefficients $C{ij}$, nine dipole--quadrupole coefficients $K_{i,\alpha}$, and six dipole--quadrupole coefficients $L_{i,\alpha}$. The total count is therefore
$3+3+3+2+9+9+6=35$.
Tracing over either subsystem recovers the corresponding single-particle density matrix above. See Appendix for details.

Two features of the $2\times3$ Hilbert space are particularly important for entanglement. First, since $\min(d_A,d_B)=2$, every pure state has Schmidt rank at most two, so its entanglement is bounded by one ebit. More importantly, the Horodecki theorem makes the positive-partial-transpose (PPT) criterion exact for $2\times3$ systems: positivity of the partial transpose is not only necessary but also sufficient for separability~\cite{Peres1996,Horodecki1996}. If the full density matrix were known, entanglement testing would therefore reduce to a direct eigenvalue test of $\rho^{T_B}$. 

Under partial transposition of the vector-meson subsystem (the operation denotes as $T_B$ as a subscript of the density matrix), the operator basis separates naturally into $T_B$-even components
${S_x,S_z,T_{zz},T_{xz},T_{xx-yy}}$
and $T_B$-odd components
${S_y,T_{yz},T_{xy}}$.
Because the $\Lambda$-side operators are unaffected, $T_B$ changes the sign of exactly 12 of the 35 parameters while leaving the remaining 23 unchanged. The question is therefore not simply whether the full $\Lambda V$ state is entangled, but whether entanglement can still be established from the experimentally accessible sector of an intrinsically incomplete tomography. Note that mixed-dimension final states have been examined in Higgs decays~\cite{BanackiEtAl2026,Barr2022Higgs,AshbyPickering2023}, but there the state is pure and tomographically complete, so the incompleteness studied here does not arise.

Specifically, the parity-violating decay $\Lambda\to p\pi^-$ is self-analyzing and provides access to all three components of the $\Lambda$ polarization, whereas the parity-conserving decay $V\to PP$ is insensitive to the vector-meson dipole polarization $P_i$~\cite{KloetTabakin}. Consequently, the three $P_i$ and nine dipole--dipole correlations $C_{ij}$ are intrinsically inaccessible, leaving 23 measurable parameters (See Table.~\ref{tab:params}). Of these, $I_{10}$, $I_{1m}$, and the six $L_{i,\alpha}$ require the full two-dimensional $(\theta,\phi)$ decay distribution and are lost in the conventional one-dimensional Schilling--Wolf projection~\cite{SchillingWolf1970}. The justification usually offered is parity conservation in production, but that argument covers only four of the eight: $I_{10}$, $I_{1m}$, $L_{n,yz}$ and $L_{n,xy}$ are parity-odd and must vanish, whereas $L_{r,\alpha}$ and $L_{k,\alpha}$ are parity-even and need not (Sec.~\ref{sec:sm_5}). Thus, complete reconstruction of the $\Lambda V$ density matrix is impossible from these decays alone, motivating the incomplete-tomography problem addressed below.

\begin{table}[t]
\centering
\begin{footnotesize}
\begin{tabular}{@{}lc@{}}
\toprule
Sector & Accessible? \\
\midrule
$a_i$ ($\Lambda$ polarization, 3) & yes, standard \\
$\rho_{00},R_{10},R_{1m}$ ($V$ alignment, 3) & yes, standard \\
$K_{i,\alpha}$ (dipole$\otimes$$V$-quad., 9) & yes, joint fit \\
$I_{10},I_{1m},L_{i,\alpha}$ ($V$ align.\ + corr., 8) & yes, 2D fit only \\
$P_i$, $C_{ij}$ ($V$ dipole $\pm$ corr., 12) & \textbf{no} \\
\bottomrule
\end{tabular}
\end{footnotesize}
\caption{Accessibility of the 35 joint parameters. Only the bottom row, 12 of 35, is beyond reach on kinematic grounds.}
\label{tab:params}
\end{table}

\textbf{\textit{Results.}} Let $\rho_{\rm meas}$ denote the density matrix obtained by setting the 23 measured parameters to their observed values and the inaccessible 12 to zero, which is the most complete reconstruction available from the decay angular distributions alone. The key observation is that this operation has a simple symmetry interpretation. Zeroing a parameter can be viewed as an averaging operation: if a linear map $\mathcal G$ reverses its sign while leaving all other parameters unchanged, then $\tfrac12[\rho+\mathcal G(\rho)]$ removes that component. Since the missing 12 parameters constitute exactly the meson dipole sector, the required transformation must reverse every $S_i$ while preserving every $T_{jk}$. This is precisely the action of time reversal on the meson spin.

The usual time-reversal operator $\Theta=UK$, where $K$ is the complex conjugation operator, is antiunitary and therefore cannot itself be applied to only one factor of a bipartite density matrix. For a Hermitian operator $O$, however, its action can be written as $\Theta O\Theta^{-1}=UO^TU^\dagger$: a transposition followed by a unitary rotation. Defining $\mathcal U_B(\rho)\equiv(\mathds{1}_2\otimes U)\rho(\mathds{1}_2\otimes U)^\dagger$, with $U=-e^{-i\pi S_y}$ corresponding to a $\pi$ rotation of the meson spin about $\hat y$, averaging over $\mathbb Z_2={\mathds{1},\mathcal U_B\circ T_B}$ gives
\begin{equation}
\rho_{\rm meas} = \tfrac12\big[\rho_{\rm true}+\mathcal U_B\big(\rho_{\rm true}^{T_B}\big)\big],
\label{eq:meas}
\end{equation}
the state averaged with its own \emph{partial time reversal}. The composite transformation $\mathcal U_B\circ T_B$ flips exactly $P_i$ and $C_{ij}$ while leaving all 23 accessible parameters invariant. Like the partial transpose operation, this transformation is positive but not completely positive and hence does not represent a physical quantum channel---a property that will be essential for its sensitivity to entanglement.

\resultbox{Lemma 1 (PPT reduces to positivity)}{For the reconstruction built from any input state, $\rho_{\rm meas}^{T_B}=\mathcal U_B(\rho_{\rm meas})$ exactly. Since $\mathcal U_B$ is a unitary conjugation, $\rho_{\rm meas}^{T_B}$ and $\rho_{\rm meas}$ share a spectrum, so the PPT test on $\rho_{\rm meas}$ reduces to asking whether $\rho_{\rm meas}\succeq0$: a single diagonalization, with no partial transpose needed in practice.}

The lemma follows directly from the transformation properties of the surviving operators: each is either invariant under both $T_B$ and $\mathcal U_B$, or changes sign under both. Thus the PPT test, which would ordinarily require constructing and diagonalizing the partial transpose, reduces here to the positivity of $\rho_{\rm meas}$ itself. We now ask the two questions relevant for an incomplete reconstruction: can the missing information generate a false entanglement signal, and conversely, when is the available information sufficient to establish entanglement?

\resultbox{Theorem 1 (No false positives)}{If $\rho_{\rm true}$ is separable, then $\rho_{\rm meas}$ is separable, for any separable $\rho_{\rm true}$, pure or mixed. A negative eigenvalue of $\rho_{\rm meas}$ can therefore never be produced by a genuinely separable state.}

\emph{Proof.} We first note that if $M$ is a legitimate single-qutrit density matrix, so is $UM^TU^\dagger$: $M^T$ is Hermitian, unit-trace, and shares the eigenvalues of $M$ (identical characteristic polynomial), hence is positive semi-definite, and conjugation by the unitary $U$ preserves Hermiticity, trace and spectrum. Now write a separable state as $\rho_{\rm true}=\sum_kp_k\,\rho^{(k)}_A\otimes\rho^{(k)}_B$ with $p_k\ge0$, $\sum_kp_k=1$. Both $T_B$ and $\mathcal U_B$ are linear and act only on the $B$ factor, so $\mathcal U_B(\rho_{\rm true}^{T_B})=\sum_kp_k\,\rho^{(k)}_A\otimes U(\rho^{(k)}_B)^TU^\dagger$, in which each meson factor is a legitimate qutrit state by the preceding remark. Substituting into Eq.~\eqref{eq:meas},
\begin{equation}
\rho_{\rm meas}=\tfrac12\sum_kp_k\,\rho^{(k)}_A\otimes\Big[\rho^{(k)}_B+U\big(\rho^{(k)}_B\big)^TU^\dagger\Big],
\end{equation}
an explicit convex combination of legitimate product states with weights $p_k/2\ge0$ summing to one. By definition this is separable, hence positive semi-definite. $\blacksquare$

Theorem~1 establishes that the reconstruction cannot manufacture an entanglement signal from a separable state. The stronger question is whether this test is also optimal given the information that is experimentally available.

\begin{figure}[t]
    \centering
    \includegraphics[width=0.48\textwidth]{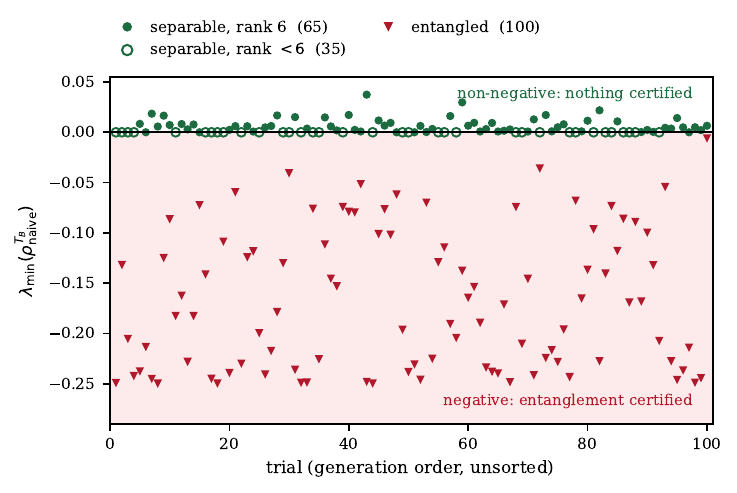}
    \caption{Numerical verification of Theorem 1. For each trial a random input state is generated, the reconstruction $\rho_{\rm meas}$ is built from its 23 accessible parameters, and the partial transpose is diagonalized. \emph{Left:} $\lambda_{\min}(\rho_{\rm meas}^{T_B})$ for 100 separable states --- 65 of full rank 6 (filled circles) and 35 of reduced rank (open circles) --- together with 100 entangled states (triangles). This numerical analysis is generated by Claude Opus 5 model.}
    \label{fig:validation}
\end{figure}

\resultbox{Theorem 2 (Exact certifiability)}{Let $\mathcal C(D)={\rho\ \text{physical}: \rho\ \text{reproduces the measured data}\ D}$. Then $\exists$ separable $\rho\in\mathcal C(D)$ if and only if $\rho_{\rm meas}$ is separable, if and only if $\rho_{\rm meas}\succeq0$. Entanglement is therefore certifiable from $D$ if and only if $\rho_{\rm meas}$ has a negative eigenvalue, and no analysis of the same data can do better.}

\emph{Proof:} By construction $\rho_{\rm meas}$ is an affine function of the measured data $D$ alone: it fixes every accessible parameter at its observed value and sets the remaining twelve to zero, and therefore does not depend on the unmeasured parameters of whichever physical state produced $D$. Every $\rho\in\mathcal C(D)$ consequently yields the same measured reconstruction, and whenever $\rho_{\rm meas}$ is a valid state it belongs to $\mathcal C(D)$.

For the first equivalence, if $\rho_{\rm meas}$ is separable then it is in particular a legitimate state, and hence a separable member of $\mathcal C(D)$. Conversely, if some separable $\rho_s\in\mathcal C(D)$ exists, then averaging $\rho_s$ as in Eq.~\eqref{eq:meas} returns $\rho_{\rm meas}$, which Theorem~1 guarantees to be separable. Existence of a separable state compatible with $D$ is thus equivalent to separability of $\rho_{\rm meas}$ itself.

For the second equivalence the relevant criterion is the Peres--Horodecki criterion, which in $2\times3$ is both necessary and sufficient: $\rho_{\rm meas}$ is separable if and only if $\rho_{\rm meas}^{T_B}\succeq0$. By Lemma~1, $\rho_{\rm meas}^{T_B}=\mathcal U_B(\rho_{\rm meas})$ is a unitary conjugation of $\rho_{\rm meas}$ and shares its spectrum, so $\rho_{\rm meas}^{T_B}\succeq0$ if and only if $\rho_{\rm meas}\succeq0$. Hence $\rho_{\rm meas}$ is separable if and only if $\rho_{\rm meas}\succeq0$.

Consequently, if $\rho_{\rm meas}$ has a negative eigenvalue then no separable state reproduces the data $D$; equivalently, entanglement is certified exactly by the negativity of $\rho_{\rm meas}$. Since the argument uses $D$ alone, no other analysis of the same data can certify more. $\blacksquare$ 

Note that the above two theorems have been independently proved by Lean Theorem Prover~\cite{LEAN,mathlib2020}.

The two results give a particularly simple interpretation of incomplete QQQT. The relevant question is not whether a particular completion of the density matrix is entangled, but whether \emph{every} physical state consistent with the measured data must be entangled~\cite{AudenaertPlenio2006,GuhneToth2009}. Theorem~2 answers this question exactly. Operationally, one constructs $\rho_{\rm meas}$ from the 23 measured moments and diagonalizes it. A negative eigenvalue implies that every state compatible with the data is entangled; if $\rho_{\rm meas}$ is positive semidefinite, at least one compatible separable state exists. The semidefinite optimization required for general incomplete tomography therefore reduces in this system to the single diagonalization of Lemma~1. Figure~\ref{fig:validation} verifies Theorem~1 numerically using randomly generated states.

Theorem~2 establishes that no analysis of the same 23 observables can outperform the reconstruction. Nevertheless, incomplete tomography can still miss genuine entanglement, and this loss can be quantified.

\resultbox{Proposition 1 (Pure states retain half the negativity)}{For a Schmidt-rank-2 state $|\psi\rangle=\sqrt{\lambda_1}|1\rangle_A|1\rangle_B+\sqrt{\lambda_2}|2\rangle_A|2\rangle_B$, in the physically relevant meson Schmidt bases, $\lambda_{\min}(\rho_{\rm meas}^{T_B})=-\tfrac12\sqrt{\lambda_1\lambda_2}$, exactly half the true value $-\sqrt{\lambda_1\lambda_2}$.}

\resultbox{Proposition 2 (Mixed-state blind windows)}{Genuine blind windows exist for mixed states. A maximally entangled state diluted with isotropic noise, $\rho(p)=p,|e_0^+\rangle\langle e_0^+|+\tfrac{1-p}{6}\mathds{1}_6$, is entangled for $p>1/4$ but uncertifiable from the 23 accessible parameters until $p>2/5$. A mixture of two maximally entangled Bell-like states is entangled for every $p\in[0,1]$ yet uncertifiable over a window of width $\sqrt2-1\approx0.41$ centered at $p=\tfrac12$.}

Thus, pure-state entanglement remains visible with a reduced negativity, whereas mixed states can contain genuine blind regions. By Theorem~2, these are fundamental limitations of the accessible observables rather than shortcomings of the reconstruction: within each blind window, at least one separable state reproduces the same measured data. Verification of Propositions are included in the Appendix.

\textbf{\textit{Remarks:}} We found the above results striking. Although this might be an coincidence, the 12 missing parameters due to the strong decay of the vector meson is somehow \textit{special}. For example, discarding
just the three dipoles $P_i$ alone, or the nine $C_{ij}$ alone, or the nine correlations $K_{i,\alpha}$, or
both single-particle polarizations $\{a_i,P_i\}$, each yields separable inputs reconstructed with a
negative eigenvalue, which are false positives; then, our written theorems would be invalid. We pushed this further by randomly picking $12$-element subsets to be zeroed, and many had failed the same way. This finding, only in this particular condition, enables the QQQT of this hadronic systems of $\Lambda\phi$ and $\Lambda K^{\ast 0}$ with incomplete density matrix. 

\textbf{\textit{Physics implications.}} Let $p(\Delta R)$ denote the fraction of spin coherence a quark retains through hadronization, one factor per hadron formed. In open-quantum-system language, where hadronization acts on the initial $s\bar s$ spin state as a quantum channel~\cite{BarataEtAl2026}, $p$ is the strength of the local map on one side; for $\Lambda\bar\Lambda$ two such maps act independently, so that $C^{\rm meas}_{ij}=p^{2}\,C^{\rm parton}_{ij}$. It runs with the pair separation because a wider separation means more string breaking in between. Measuring both systems then gives the following.
\begin{itemize}
\setlength{\itemsep}{2pt}\setlength{\parskip}{0pt}

\item \emph{Constraints to quark-level correlation.} Each hadron is dressed locally, and local dressing cannot create entanglement, so whatever spin correlation survives into the observed hadrons is a lower limit on what the quark pair carried at production, independently of how hadronization works. This can be used to distinguish initial-state verse final-state effects~\cite{Zhang:2026tqs} that provide spin correlations.

\item \emph{Baryon--meson comparison.} A $\Lambda$ forms when the $s$ picks up a $ud$ diquark, whereas a vector meson forms when the $\bar s$ picks up a single partner; if coherence is lost by entangling with the surrounding string, dragging along two extra quarks need not cost the same as dragging along one. The comparison isolates this, although not through the same observable in the two systems: for $\Lambda\bar\Lambda$ the coherence is read directly off $C_{ij}$, whereas for $\Lambda V$ that dipole--dipole sector is precisely what is lost, and the meson factor enters only through $K_{i,\alpha}$ and $L_{i,\alpha}$. Comparing how fast the two systems degrade, rather than their absolute normalizations, is what separates $p_{\rm meson}$ from $p_{\rm baryon}$, a new physics quantity chracterizing hadronization that has never been measured.

\item \emph{Quantum-to-classical transitions.} Another physics question we can ask is how far out in $\Delta R$ each system remains quantum entangled. The $\Lambda\bar\Lambda$ system exposes the full spin state, while $\Lambda V$ necessarily reports less than it holds: the reconstruction can miss entanglement but, by Theorem~1, can never create it. Consequently, if $\Lambda V$ is still certified at separations where $\Lambda\bar\Lambda$ has gone classical, the conclusion follows immediately: spin coherence survives better inside a meson than inside a baryon.
\end{itemize}

\textbf{\textit{Summary.}} We have developed a qubit--qutrit quantum tomography (QQQT) technique for spin-$\tfrac12\otimes1$ $\Lambda V$ systems and shown that, despite the intrinsically \textit{incomplete} reconstruction, the 23 experimentally accessible spin parameters are sufficient to provide an exact criterion for when entanglement is implied by the data. The resulting test reduces the general incomplete-tomography problem to a single diagonalization, while its loss of sensitivity can be quantified analytically for both pure and mixed states. Applied to $\Lambda\phi$ and $\Lambda K^{*0}$ production, this framework requires only angular-moment measurements in existing data sets and provides a new complement to $\Lambda\bar\Lambda$ studies~\cite{STARNature2026}: comparing baryon--meson and baryon--antibaryon spin correlations can directly probe how quantum coherence evolves through different hadronization channels~\cite{LiuTu2026,BarataEtAl2026,DattaEtAl2025}. More broadly, the same construction applies to any spin-$\tfrac12\otimes1$ system analyzed through a self-analyzing weak decay and a parity-conserving two-body decay.

\section*{Acknowledgments.}
The authors thank the Department of Energy's Summer Undergraduate Laboratory Internship (SULI) for supporting V.Kanachova's research project at Brookhaven National Lab. We also acknowledge useful informal discussions with members of the local ePIC group at Brookhaven National Laboratory. The authors acknowledge the usage of Artificial Intelligence models, ChatGPT 5 and Claude Opus 5, to polish the language as well as perform numerical verifications.  The work of F. Liu and Z. Tu was supported by the U.S. Department of Energy under Award No. DE-SC0012704 and by the Brookhaven National Laboratory Laboratory Directed Research and Development (LDRD) Project 26-029.

%

\setcounter{equation}{0}
\renewcommand{\theequation}{S\arabic{equation}}
\makeatletter
\@ifundefined{theHequation}{}{\renewcommand{\theHequation}{SM.\arabic{equation}}}
\makeatother
\setcounter{table}{0}
\renewcommand{\thetable}{S\arabic{table}}
\makeatletter
\@ifundefined{theHtable}{}{\renewcommand{\theHtable}{SM.\arabic{table}}}
\makeatother

\begin{center}
{\large\bf Appendix}
\end{center}

\noindent
The Appendix is organized as follows. Sections~\ref{sec:sm_1}--\ref{sec:sm_2} fix the state and the frame. Sections~\ref{sec:sm_3}--\ref{sec:sm_5} determine experimental accessibility: we construct the two decay analyzers, evaluate the master angular distribution $W(\theta_1,\phi_1,\theta_2,\phi_2)$ in closed form, read the $23/12$ accessibility split directly from it, give an estimator for each accessible parameter, and classify the parameters by parity. Section~\ref{sec:sm_LL} provides a comparison with the $\Lambda\bar\Lambda$ system, where all 15 parameters are measured. Section~\ref{sec:sm_6} constructs $\rho_{\rm meas}$ and gives the proof of Lemma~1 and the verification of Propositions~1 and~2.

\subsection{Conventions: frame, operators, and the joint state\label{sec:sm_1}}
All spin components are defined with respect to the pair-rest-frame triad $\{\hat k,\hat r,\hat n\}$ of Fig.~\ref{fig:helicity}: $\hat k$ along the $\Lambda$ momentum, $\hat n\propto\hat p_{\rm beam}\times\hat k$, and $\hat r=\hat n\times\hat k$, so that $\hat k\times\hat r=\hat n$. Throughout we identify
\begin{equation}
(\hat x,\hat y,\hat z)\;\equiv\;(\hat r,\hat n,\hat k),
\label{eq:S:axes}
\end{equation}
which makes $(\hat x,\hat y,\hat z)$ right-handed and fixes the index convention throughout.

The Pauli matrices are $\sigma_x=\left(\begin{smallmatrix}0&1\\1&0\end{smallmatrix}\right)$, $\sigma_y=\left(\begin{smallmatrix}0&-i\\i&0\end{smallmatrix}\right)$, $\sigma_z=\left(\begin{smallmatrix}1&0\\0&-1\end{smallmatrix}\right)$, and the spin-1 generators in the helicity basis $\{|{+1}\rangle,|0\rangle,|{-1}\rangle\}$ are
\[
S_z=\begin{pmatrix}1&0&0\\0&0&0\\0&0&-1\end{pmatrix},\quad
S_x=\tfrac1{\sqrt2}\begin{pmatrix}0&1&0\\1&0&1\\0&1&0\end{pmatrix},
\]
\[
S_y=\tfrac1{\sqrt2}\begin{pmatrix}0&-i&0\\i&0&-i\\0&i&0\end{pmatrix}.
\]
The quadrupole tensor is $T_{jk}\equiv\tfrac32(S_jS_k+S_kS_j)-2\delta_{jk}\mathds{1}_3$, with five independent components $T_{zz},T_{xx-yy},T_{xy},T_{xz},T_{yz}$; e.g.\ $T_{zz}=\mathrm{diag}(1,-2,1)$, and all satisfy $\mathrm{Tr}\,T_{jk}=0$. A fact used repeatedly below is that $S_x,S_z,T_{zz},T_{xz},T_{xx-yy}$ are real, whereas $S_y,T_{yz},T_{xy}$ are purely imaginary; this underlies the sign-flip rule of Sec.~\ref{sec:sm_6}.

In the operator basis $\{\mathds{1}_2,\sigma_i\}\otimes\{\mathds{1}_3,S_i,T_{jk}\}$ the joint state is Eq.~\eqref{eq:rhofull} of the main text,
\begin{align}
\rho={}&\tfrac16\Big[\mathds{1}_2\!\otimes\!\mathds{1}_3+a_i(\sigma_i\!\otimes\!\mathds{1}_3)+\mathds{1}_2\!\otimes\!\big(\rho_{00}T_{zz}+R_{10}T_{xz}\notag\\
&+R_{1m}T_{xx-yy}\big)+\mathds{1}_2\!\otimes\!(P_xS_x+P_yS_y+P_zS_z)\notag\\
&+\mathds{1}_2\!\otimes\!(I_{10}T_{yz}+I_{1m}T_{xy})+C_{ij}(\sigma_i\!\otimes\!S_j)\notag\\
&+K_{i,\alpha}(\sigma_i\!\otimes\!T_\alpha^{\rm even})+L_{i,\alpha}(\sigma_i\!\otimes\!T_\alpha^{\rm odd})\Big],
\label{eq:S:rhofull}
\end{align}
with $\alpha^{\rm even}\in\{zz,xz,xx\!-\!yy\}$, $\alpha^{\rm odd}\in\{yz,xy\}$ and the count $3+3+3+2+9+9+6=35$.

One convention matters when comparing with the spin-alignment literature: tracing out the meson reproduces $\rho_A$ exactly, but tracing out the $\Lambda$ returns the operator content of $\rho_B$ with \emph{rescaled} coefficients, $P_i^{(3)}=\tfrac23P_i$ and quadrupole$^{(3)}=\tfrac13\times$quadrupole relative to the single-particle convention of Eq.~(1). All values below use the joint convention of Eq.~\eqref{eq:S:rhofull}.

\subsection{Basis states\label{sec:sm_2}}
Two bases recur below: the Bell-like states label the mixed-state families, while the Clebsch--Gordan states provide the physically motivated Schmidt bases used in Sec.~\ref{sec:sm_6}.

\emph{Product basis} --- 6 states $|a,b\rangle$, $a\in\{\uparrow,\downarrow\}$, $b\in\{+1,0,-1\}$.

\emph{Generalized Bell basis} --- since $\min(d_A,d_B)=2$, entanglement is capped at 1 ebit (Schmidt rank $\le2$); a complete orthonormal basis of maximally entangled states is
\begin{equation}
|e_b^\pm\rangle=\tfrac{1}{\sqrt2}\big(|{\uparrow},b\rangle\pm|{\downarrow},b{+}1\rangle\big),\quad b=+1,0,-1\ (\mathrm{mod}\ 3),
\label{eq:S:bell}
\end{equation}
indexed $e_0,e_1,e_2$ for $b=+1,0,-1$, so $|e_0^+\rangle=\tfrac1{\sqrt2}(|{\uparrow},{+1}\rangle+|{\downarrow},0\rangle)$ and $|e_1^+\rangle=\tfrac1{\sqrt2}(|{\uparrow},0\rangle+|{\downarrow},{-1}\rangle)$.

\emph{Clebsch--Gordan basis} --- from $\tfrac12\otimes1\to J=\tfrac32\oplus\tfrac12$:
\begin{align}
\big|\tfrac32,\tfrac12\big\rangle&=\sqrt{\tfrac23}|{\uparrow},0\rangle+\sqrt{\tfrac13}|{\downarrow},{+1}\rangle,\\
\big|\tfrac12,\tfrac12\big\rangle&=\sqrt{\tfrac13}|{\uparrow},0\rangle-\sqrt{\tfrac23}|{\downarrow},{+1}\rangle,
\end{align}
(and the $M\to-M$ images), entangled with Schmidt coefficients $(\tfrac13,\tfrac23)$, not maximal; the stretched states $|\tfrac32,\pm\tfrac32\rangle$ are product states.

\subsection{Decay analyzers and the master angular distribution\label{sec:sm_3}}
All experimentally accessible information in Eq.~\eqref{eq:S:rhofull} is contained in a single function of four angles. Let $\hat q_1$ be the proton direction from $\Lambda\to p\pi^-$, evaluated in the $\Lambda$ rest frame, and $\hat q_2$ the $K^+$ direction from $V\to PP$, evaluated in the $V$ rest frame, both resolved on the common axes of Eq.~\eqref{eq:S:axes}, with polar and azimuthal angles $(\theta_1,\phi_1)$ and $(\theta_2,\phi_2)$ as in Fig.~\ref{fig:helicity}. Write
\begin{equation}
\hat n(\theta_1,\phi_1)=(\sin\theta_1\cos\phi_1,\ \sin\theta_1\sin\phi_1,\ \cos\theta_1)
\label{eq:S:nhat}
\end{equation}
for the proton unit vector.

The two decays have very different analyzers. The weak, parity-violating $\Lambda\to p\pi^-$ has the fully informative analyzer
\begin{equation}
F^\Lambda(\theta_1,\phi_1)=\mathds{1}_2+\alpha_\Lambda\,\hat n(\theta_1,\phi_1)\cdot\boldsymbol\sigma,\qquad \alpha_\Lambda\approx0.75,
\label{eq:S:Flam}
\end{equation}
which contains the rank-0 and full rank-1 (dipole) content of the $\Lambda$ spin. The strong/electromagnetic, parity-conserving $V\to PP$ proceeds through the single helicity amplitude $V(0,0)$, so its analyzer $G^V(\theta_2,\phi_2)$ is built from rank-0 and rank-2 multipoles \emph{only}~\cite{AjaltouniConte,KloetTabakin}: a two-body decay into two spinless particles couples to $L=0,2$ and never to $L=1$, irrespective of how the meson was produced. This structural asymmetry between $F^\Lambda$ and $G^V$ determines the accessibility results below.

Evaluating $W\propto\mathrm{Tr}\big[\rho\,(F^\Lambda\otimes G^V)\big]$ for the full 35-parameter $\rho$ gives a factorized result in which the correlation sector uses the \emph{same} five meson-side angular functions as the single-particle sector:
\begin{align}
W(\theta_1,\phi_1,\theta_2,\phi_2)\;\propto\;{}& 1\notag\\[2pt]
&+\,3\big[\rho_{00}\,g_{00}+R_{10}\,g_{10}+R_{1m}\,g_{1m}\notag\\
&\qquad\ +I_{10}\,h_{10}+I_{1m}\,h_{1m}\big]\notag\\[2pt]
&+\,\alpha_\Lambda\,\hat n(\theta_1,\phi_1)\cdot a\notag\\[2pt]
&+\,3\alpha_\Lambda\!\!\sum_{i=x,y,z}\!\!\hat n_i(\theta_1,\phi_1)\Big[K_{i,zz}\,g_{00}\notag\\
&\qquad+K_{i,xz}\,g_{10}+K_{i,xx\text{-}yy}\,g_{1m}\notag\\
&\qquad+L_{i,yz}\,h_{10}+L_{i,xy}\,h_{1m}\Big],
\label{eq:S:Wmaster}
\end{align}
The four blocks are the isotropic term, meson alignment (5 terms, $V$ angles only), $\Lambda$ polarization (3 terms, $\Lambda$ angles only), and correlations ($3\times5=15$ genuine four-angle terms). Here every $g_\alpha,h_\alpha$ is evaluated at $(\theta_2,\phi_2)$. The five meson-side functions are
\begin{equation}
\begin{aligned}
g_{00}(\theta_2,\phi_2)&=\sin^2\theta_2-\tfrac23, \\
g_{10}(\theta_2,\phi_2)&=-\tfrac12\sin2\theta_2\cos\phi_2,\\
g_{1m}(\theta_2,\phi_2)&=-\sin^2\theta_2\cos2\phi_2,\\
h_{10}(\theta_2,\phi_2)&=\tfrac12\sin2\theta_2\sin\phi_2,\\
h_{1m}(\theta_2,\phi_2)&=\tfrac12\sin^2\theta_2\sin2\phi_2 .
\end{aligned}
\label{eq:S:harmonics}
\end{equation}
The relative block factors $(1,\,3,\,\alpha_\Lambda,\,3\alpha_\Lambda)$ follow from the exact trace evaluation and have been verified numerically. Setting $I_{10}=I_{1m}=0$ and dropping the correlation block reduces Eq.~\eqref{eq:S:Wmaster} to the conventional three-parameter Schilling--Wolf distribution in $\rho_{00},\mathrm{Re}\,\rho_{1,0},\mathrm{Re}\,\rho_{1,-1}$ used in standard spin-alignment analyses~\cite{SchillingWolf1970,STARSpinAlignment}; the $h_\alpha$ terms are present in the decay but are not fitted there, as discussed in Sec.~\ref{sec:sm_5}.

\subsection{Accessibility: reading the 23/12 split off $W$\label{sec:sm_4}}
Equation~\eqref{eq:S:Wmaster} determines accessibility by inspection, without further dynamical input. Counting the independent angular structures it contains,
\begin{equation}
\underbrace{3}_{\hat n_i}\;+\;\underbrace{5}_{g_\alpha,h_\alpha}\;+\;\underbrace{15}_{\hat n_i\times(g_\alpha,h_\alpha)}\;=\;23 ,
\end{equation}
one for each accessible parameter: the three $a_i$, the five meson multipoles $\rho_{00},R_{10},R_{1m},I_{10},I_{1m}$, and the fifteen correlations $K_{i,\alpha}$ (nine) and $L_{i,\alpha}$ (six).

The remaining twelve parameters in Eq.~\eqref{eq:S:rhofull} --- the meson dipole $P_i$ and the nine dipole--dipole correlations $C_{ij}$ --- are \emph{identically absent} from Eq.~\eqref{eq:S:Wmaster}, for all $(\theta_1,\phi_1,\theta_2,\phi_2)$. The reason is structural: detecting $P_i$ would require a rank-1 harmonic of $(\theta_2,\phi_2)$ in $W$, and detecting $C_{ij}$ would require the product of $\hat n_i$ with such a harmonic. No rank-1 function of the meson angles occurs anywhere in Eq.~\eqref{eq:S:Wmaster}, because $G^V$ contains no $L=1$ multipole. The inaccessible sector is therefore exactly
\begin{equation}
\underbrace{3}_{P_i}\;+\;\underbrace{9}_{C_{ij}}\;=\;12 ,
\end{equation}
reproducing Table~\ref{tab:params} of the main text. Equivalently, one may add a multiple of the meson dipole operator to $\rho$, varying $P_i$ and $C_{ij}$ while leaving every moment of Eq.~\eqref{eq:S:Wmaster} unchanged. The incompleteness is therefore intrinsic, not a fit artifact.

\subsection{Method of moments: the estimator for each parameter\label{sec:sm_5}}
All 23 accessible parameters are angular moments of the single-particle or joint decay distributions of the $\Lambda\phi$ and $\Lambda K^{\ast0}$ systems and can therefore be extracted from existing data sets. The main experimental complication is that the pair acceptance in $\Omega\equiv(\theta_1,\phi_1,\theta_2,\phi_2)$ does not factorize between the two decays. This can be corrected by event mixing: a $\Lambda$ and a $V$ from different events in the same event class are paired to reproduce the detector acceptance while removing the physical two-particle correlation,
\begin{equation}
W(\Omega)\;\propto\;W_{\rm same}(\Omega)\big/W_{\rm mix}(\Omega),
\label{eq:mixed}
\end{equation}
after which the 15 angular correlations are extracted. Here event mixing serves only as an acceptance correction. Operationally, one reconstructs the $\Lambda$ and $V$ candidates, corrects the joint and single-particle angular distributions, extracts the 23 moments below, constructs $\rho_{\rm meas}$, and diagonalizes it. Carrying this out differentially in the pair separation $\Delta R$ turns the entanglement measurement into a probe of hadronization, allowing the evolution of the signal to be compared among $\Lambda\phi$, $\Lambda K^{*0}$, and control channels such as $\Lambda\rho^0$, which shares no valence quark with the $\Lambda$~\cite{PDG2024}.

The four $\Lambda$-side functions $1,\hat n_x,\hat n_y,\hat n_z$ are mutually orthogonal over the $\Lambda$ solid angle, and the six meson-side functions $1,g_{00},g_{10},g_{1m},h_{10},h_{1m}$ are mutually orthogonal over the $V$ solid angle --- all cross-integrals vanish by direct integration. Thus the 23 structures of Eq.~\eqref{eq:S:Wmaster} do not mix, and each parameter follows from a \emph{single} angular moment, without fitting. Writing $\langle f\rangle$ for the event average of $f(\theta_1,\phi_1,\theta_2,\phi_2)$ over the acceptance-corrected sample of Eq.~\eqref{eq:mixed},
\begin{align}
a_i&=\tfrac3{\alpha_\Lambda}\langle\hat n_i\rangle,\notag\\
\rho_{00}&=\tfrac{15}4\langle g_{00}\rangle,\quad R_{10}=5\langle g_{10}\rangle,\quad R_{1m}=\tfrac54\langle g_{1m}\rangle,\notag\\
I_{10}&=5\langle h_{10}\rangle,\quad I_{1m}=5\langle h_{1m}\rangle,\notag\\
K_{i,zz}&=\tfrac{45}{4\alpha_\Lambda}\langle\hat n_ig_{00}\rangle,\quad K_{i,xz}=\tfrac{15}{\alpha_\Lambda}\langle\hat n_ig_{10}\rangle,\notag\\
K_{i,xx\text{-}yy}&=\tfrac{15}{4\alpha_\Lambda}\langle\hat n_ig_{1m}\rangle,\notag\\
L_{i,yz}&=\tfrac{15}{\alpha_\Lambda}\langle\hat n_ih_{10}\rangle,\quad L_{i,xy}=\tfrac{15}{\alpha_\Lambda}\langle\hat n_ih_{1m}\rangle .
\label{eq:S:moments}
\end{align}
The prefactors combine the normalization integrals $\langle\hat n_i^2\rangle=\tfrac13$, $\langle g_{00}^2\rangle=\tfrac4{45}$, $\langle g_{10}^2\rangle=\langle h_{10}^2\rangle=\langle h_{1m}^2\rangle=\tfrac1{15}$ and $\langle g_{1m}^2\rangle=\tfrac4{15}$ with the block factors of Eq.~\eqref{eq:S:Wmaster}. Table~\ref{tab:moments} summarizes the correspondence parameter\,$\leftrightarrow$\,angular function\,$\leftrightarrow$\,estimator, written explicitly in $(\theta_1,\phi_1,\theta_2,\phi_2)$.

\begin{table*}[t]
\centering
\small
\begin{tabular}{@{}llll@{}}
\toprule
Sector & Parameter & Angular function of $(\theta_1,\phi_1,\theta_2,\phi_2)$ & Estimator \\
\midrule
\multicolumn{4}{@{}l}{\emph{$\Lambda$ polarization --- 3 parameters, $\Lambda$ angles only}}\\
 & $a_x$ & $\hat n_x=\sin\theta_1\cos\phi_1$ & $\tfrac{3}{\alpha_\Lambda}\langle\hat n_x\rangle$\\
 & $a_y$ & $\hat n_y=\sin\theta_1\sin\phi_1$ & $\tfrac{3}{\alpha_\Lambda}\langle\hat n_y\rangle$\\
 & $a_z$ & $\hat n_z=\cos\theta_1$ & $\tfrac{3}{\alpha_\Lambda}\langle\hat n_z\rangle$\\
\midrule
\multicolumn{4}{@{}l}{\emph{meson alignment --- 5 parameters, $V$ angles only}}\\
 & $\rho_{00}$ & $g_{00}=\sin^2\theta_2-\tfrac23$ & $\tfrac{15}{4}\langle g_{00}\rangle$\\
 & $R_{10}$ & $g_{10}=-\tfrac12\sin2\theta_2\cos\phi_2$ & $5\langle g_{10}\rangle$\\
 & $R_{1m}$ & $g_{1m}=-\sin^2\theta_2\cos2\phi_2$ & $\tfrac54\langle g_{1m}\rangle$\\
 & $I_{10}$ & $h_{10}=\tfrac12\sin2\theta_2\sin\phi_2$ & $5\langle h_{10}\rangle$\\
 & $I_{1m}$ & $h_{1m}=\tfrac12\sin^2\theta_2\sin2\phi_2$ & $5\langle h_{1m}\rangle$\\
\midrule
\multicolumn{4}{@{}l}{\emph{correlations --- 15 parameters, genuine four-angle moments ($i=x,y,z$)}}\\
 & $K_{i,zz}$ & $\hat n_i(\theta_1,\phi_1)\,\big(\sin^2\theta_2-\tfrac23\big)$ & $\tfrac{45}{4\alpha_\Lambda}\langle\hat n_i\,g_{00}\rangle$\\
 & $K_{i,xz}$ & $-\tfrac12\,\hat n_i(\theta_1,\phi_1)\,\sin2\theta_2\cos\phi_2$ & $\tfrac{15}{\alpha_\Lambda}\langle\hat n_i\,g_{10}\rangle$\\
 & $K_{i,xx\text{-}yy}$ & $-\,\hat n_i(\theta_1,\phi_1)\,\sin^2\theta_2\cos2\phi_2$ & $\tfrac{15}{4\alpha_\Lambda}\langle\hat n_i\,g_{1m}\rangle$\\
 & $L_{i,yz}$ & $\tfrac12\,\hat n_i(\theta_1,\phi_1)\,\sin2\theta_2\sin\phi_2$ & $\tfrac{15}{\alpha_\Lambda}\langle\hat n_i\,h_{10}\rangle$\\
 & $L_{i,xy}$ & $\tfrac12\,\hat n_i(\theta_1,\phi_1)\,\sin^2\theta_2\sin2\phi_2$ & $\tfrac{15}{\alpha_\Lambda}\langle\hat n_i\,h_{1m}\rangle$\\
\midrule
\multicolumn{4}{@{}l}{\emph{inaccessible --- 12 parameters: no rank-1 harmonic of $(\theta_2,\phi_2)$ exists in $W$}}\\
 & $P_i$ & would require $Y_{1m}(\theta_2,\phi_2)$ & absent from $W$\\
 & $C_{ij}$ & would require $\hat n_i(\theta_1,\phi_1)\,Y_{1m}(\theta_2,\phi_2)$ & absent from $W$\\
\bottomrule
\end{tabular}
\caption{Correspondence between the parameters of Eq.~\eqref{eq:S:rhofull}, the orthogonal angular structures of Eq.~\eqref{eq:S:Wmaster}, and their moment estimators. The 23 accessible parameters split as $3+5+15$; the 12 inaccessible ones are exactly those that would need a rank-1 (dipole) harmonic of the meson decay angles, which $V\to PP$ does not provide. The eight parameters $I_{10},I_{1m},L_{i,yz},L_{i,xy}$ are carried entirely by the $\sin\phi_2$ and $\sin2\phi_2$ structures omitted by the conventional Schilling--Wolf fit.}
\label{tab:moments}
\end{table*}

\medskip\noindent\emph{Parity classification.}\ \ Parity conservation in production predicts that roughly half of the 23 accessible parameters vanish, providing internal null tests. In the pair rest frame momenta reverse under parity while $\hat n\propto\hat p_{\rm beam}\times\hat k$, a cross product of two polar vectors, is invariant; spin is axial, so a spin \emph{component} picks up the sign of its axis, $s_r=-1$, $s_n=+1$, $s_k=-1$. A term survives iff the product of $s$ over its indices is $+1$, equivalently, iff it carries an odd number of $\hat n$ labels. Applying this rule, 11 of the 23 are parity-allowed, namely $a_n$; $\rho_{00},R_{10},R_{1m}$; the three $K_{n,\alpha}$; and the four $L_{r,\alpha},L_{k,\alpha}$; while the remaining 12, $a_r,a_k$; $I_{10},I_{1m}$; the six $K_{r,\alpha},K_{k,\alpha}$; and $L_{n,yz},L_{n,xy}$, vanish if production conserves parity. The first group includes the familiar results that global hyperon polarization lies along the production-plane normal and that only $\rho_{00},\mathrm{Re}\,\rho_{1,0},\mathrm{Re}\,\rho_{1,-1}$ are conventionally fitted.

The constraint applies to production, not to the decays: the parity-violating $\Lambda\to p\pi^-$ is what makes the hyperon self-analyzing and is already carried by $F^\Lambda$ in Eq.~\eqref{eq:S:Flam}. Since production is strong and electromagnetic, the 12 parity-odd parameters are expected to vanish and therefore provide a cross-check of the acceptance correction in Eq.~\eqref{eq:mixed}: detector holes do not respect the mirror symmetry, so an imperfectly corrected acceptance leaks into precisely these terms while leaving the 11 parity-allowed ones comparatively undisturbed. A significant nonzero value therefore points first to the acceptance or the orientation convention for $\hat n$, and only then to background or genuinely parity-violating production. This does not affect the entanglement analysis, which uses all 23 measured values.

\subsection{Complete tomography of the $\Lambda\bar\Lambda$ system\label{sec:sm_LL}}
For comparison with the $\Lambda V$ case, the same programme is complete for the $\Lambda\bar\Lambda$ system. Both members of the pair are spin-$\tfrac12$, so the joint Hilbert space is $2\times2$ and, in the operator basis $\{\mathds{1}_2,\sigma_i\}^{\otimes2}$, a Hermitian unit-trace density matrix carries $16-1=15$ independent real parameters,
\begin{equation}
\rho_{\Lambda\bar\Lambda}=\tfrac14\Big[\mathds{1}_2\!\otimes\!\mathds{1}_2+B^{+}_i\,(\sigma_i\!\otimes\!\mathds{1}_2)+B^{-}_j\,(\mathds{1}_2\!\otimes\!\sigma_j)+C_{ij}\,(\sigma_i\!\otimes\!\sigma_j)\Big],
\label{eq:S:rhoLL}
\end{equation}
counted as $\underbrace{3}_{B^{+}_i}+\underbrace{3}_{B^{-}_j}+\underbrace{9}_{C_{ij}}=15$, with $B^{\pm}$ the single-hyperon polarization vectors and $C_{ij}=\langle\sigma_i\otimes\sigma_j\rangle$ the spin-spin correlation matrix. All components are referred to the same triad $(\hat x,\hat y,\hat z)\equiv(\hat r,\hat n,\hat k)$ of Sec.~\ref{sec:sm_1}, with $\hat k$ now along the $\Lambda$ momentum in the $\Lambda\bar\Lambda$ rest frame.

The key difference from Sec.~\ref{sec:sm_3} is that \emph{both} analyzers are fully informative. Each hyperon decays weakly, $\Lambda\to p\pi^-$ and $\bar\Lambda\to\bar p\pi^+$, so each carries a nonvanishing rank-1 analyzing power along all three axes,
\begin{equation}
F^{\Lambda}=\mathds{1}_2+\alpha_\Lambda\,\hat q_1\!\cdot\!\boldsymbol\sigma,\qquad
F^{\bar\Lambda}=\mathds{1}_2+\alpha_{\bar\Lambda}\,\hat q_2\!\cdot\!\boldsymbol\sigma,
\label{eq:S:FLL}
\end{equation}
with $\alpha_{\bar\Lambda}\simeq-\alpha_\Lambda$ if $CP$ is conserved. Evaluating $W\propto\mathrm{Tr}[\rho_{\Lambda\bar\Lambda}(F^{\Lambda}\otimes F^{\bar\Lambda})]$ gives the joint decay distribution
\begin{equation}
W(\hat q_1,\hat q_2)\;\propto\;1+\alpha_\Lambda\,\hat q_1\!\cdot\!\mathbf B^{+}+\alpha_{\bar\Lambda}\,\hat q_2\!\cdot\!\mathbf B^{-}+\alpha_\Lambda\alpha_{\bar\Lambda}\,C_{ij}\,\hat q_{1i}\hat q_{2j},
\label{eq:S:WLL}
\end{equation}
where $\hat q_1$ is the proton direction evaluated in the $\Lambda$ rest frame and $\hat q_2$ the antiproton direction in the $\bar\Lambda$ rest frame, both resolved on the common axes and parametrized by $(\theta_1,\phi_1)$ and $(\theta_2,\phi_2)$ exactly as in Fig.~\ref{fig:helicity}.

Every one of the 15 parameters appears in Eq.~\eqref{eq:S:WLL}, and the angular structures $\{\hat q_{1i}\}$, $\{\hat q_{2j}\}$ and $\{\hat q_{1i}\hat q_{2j}\}$ are mutually orthogonal over the two solid angles. Each parameter is therefore isolated by a single moment, without fitting:
\begin{equation}
B^{+}_i=\frac{3}{\alpha_\Lambda}\big\langle \hat q_{1i}\big\rangle,\qquad\\
B^{-}_j=\frac{3}{\alpha_{\bar\Lambda}}\big\langle \hat q_{2j}\big\rangle,\qquad \\
C_{ij}=\frac{9}{\alpha_\Lambda\alpha_{\bar\Lambda}}\big\langle \hat q_{1i}\hat q_{2j}\big\rangle,
\label{eq:S:momentsLL}
\end{equation}
the prefactors following from $\langle\hat q_{1i}^2\rangle=\langle\hat q_{2j}^2\rangle=\tfrac13$. Written out in the four angles, $\hat q_1=(\sin\theta_1\cos\phi_1,\sin\theta_1\sin\phi_1,\cos\theta_1)$ and likewise for $\hat q_2$, so that for example $C_{rn}=9\langle\sin\theta_1\cos\phi_1\,\sin\theta_2\sin\phi_2\rangle/(\alpha_\Lambda\alpha_{\bar\Lambda})$; Table~\ref{tab:momentsLL} summarizes the full correspondence. We verified Eq.~\eqref{eq:S:momentsLL} with a Monte Carlo closure test: events generated from Eq.~\eqref{eq:S:WLL} for a random physical $\rho_{\Lambda\bar\Lambda}$ recover all 15 input parameters.

The contrast with Sec.~\ref{sec:sm_4} is immediate. For $\Lambda\bar\Lambda$ the tomography is \emph{complete}: 15 of 15 parameters are measured, $\rho_{\Lambda\bar\Lambda}$ is reconstructed in full, and the Peres--Horodecki criterion --- exact in $2\times2$ as it is in $2\times3$ --- can be applied directly to the measured matrix. No analogue of the reconstruction in Sec.~\ref{sec:sm_6} is required, and no blind windows arise. Thus the incompleteness arises specifically when one weakly decaying hyperon is replaced by a strongly decaying vector meson: it is the loss of the meson rank-1 analyzing power, not the increase in Hilbert-space dimension, that removes 12 of the 35 parameters.

\begin{table}[t]
\centering
\begin{footnotesize}
\begin{tabular}{@{}llll@{}}
\toprule
Sector & Parameter & Angular function & Estimator \\
\midrule
$\Lambda$ polarization & $B^{+}_i$ & $\hat q_{1i}(\theta_1,\phi_1)$ & $\tfrac{3}{\alpha_\Lambda}\langle\hat q_{1i}\rangle$\\
$\bar\Lambda$ polarization & $B^{-}_j$ & $\hat q_{2j}(\theta_2,\phi_2)$ & $\tfrac{3}{\alpha_{\bar\Lambda}}\langle\hat q_{2j}\rangle$\\
spin--spin correlation & $C_{ij}$ & $\hat q_{1i}\,\hat q_{2j}$ & $\tfrac{9}{\alpha_\Lambda\alpha_{\bar\Lambda}}\langle\hat q_{1i}\hat q_{2j}\rangle$\\
\midrule
\multicolumn{4}{@{}l}{total $3+3+9=15$, all accessible ($i,j=r,n,k$)}\\
\bottomrule
\end{tabular}
\end{footnotesize}
\caption{Correspondence between the 15 parameters of the $\Lambda\bar\Lambda$ density matrix Eq.~\eqref{eq:S:rhoLL}, the orthogonal angular structures of Eq.~\eqref{eq:S:WLL}, and their moment estimators. Unlike the $\Lambda V$ case of Table~\ref{tab:moments}, no sector is kinematically inaccessible: both decays are weak and self-analyzing, so each supplies a full rank-1 analyzing power.}
\label{tab:momentsLL}
\end{table}

\subsection{Reconstruction of $\rho_{\rm meas}$ and proofs\label{sec:sm_6}}
\noindent\emph{The reconstruction as a partial time reversal.}\ \ Having identified the measured sector, we turn to the entanglement test. Throughout we use the partial transpose on the meson, $\rho^{T_B}_{(a,b),(a',b')}=\rho_{(a,b'),(a',b)}$, for which the Horodecki theorem makes PPT necessary and sufficient for separability in $2\times3$~\cite{Peres1996,Horodecki1996,Horodecki1997}, and the labels $T$-even for $\{S_x,S_z,T_{zz},T_{xz},T_{xx-yy}\}$ and $T$-odd for $\{S_y,T_{yz},T_{xy}\}$, following the reality properties of Sec.~\ref{sec:sm_1}.

Setting a parameter to zero can be viewed as an averaging operation: if a linear map $\mathcal G$ flips its sign while preserving the others, then $\tfrac12[\rho+\mathcal G(\rho)]$ sets it to zero. By Sec.~\ref{sec:sm_4}, the required map flips every $S_i$ while preserving every $T_{jk}$ --- precisely time reversal on the meson factor. The textbook operator is $\Theta=UK$, with $K$ complex conjugation in the helicity basis ($\Theta|m\rangle=(-1)^m|{-m}\rangle$) and
\begin{equation}
U=\begin{pmatrix}0&0&-1\\0&1&0\\-1&0&0\end{pmatrix}\quad(\text{basis order}\ |{+1}\rangle,|0\rangle,|{-1}\rangle).
\label{eq:S:U}
\end{equation}
It acts as required, $\Theta S_i\Theta^{-1}=-S_i$ and $\Theta T_{jk}\Theta^{-1}=T_{jk}$. Because it is antiunitary, it admits no one-sided action on a tensor product; however, for Hermitian $O$ one has $O^T=O^*$ and hence $\Theta O\Theta^{-1}=UO^TU^\dagger$: time reversal is transposition \emph{followed by} a unitary rotation. Partial transposition supplies the first step; the second is
\begin{equation}
\mathcal U_B(\rho)\equiv(\mathds{1}_2\otimes U)\,\rho\,(\mathds{1}_2\otimes U)^\dagger,
\label{eq:S:UB}
\end{equation}
a rotation by $\pi$ about $\hat y$ of the meson spin ($U=-e^{-i\pi S_y}$). Its action, together with that of transposition, on the operator basis is
\begin{equation}
\begin{array}{lcccccccc}
 & S_x & S_y & S_z & T_{zz} & T_{xz} & T_{xx-yy} & T_{yz} & T_{xy}\\
O\!\mapsto\!O^T: & + & - & + & + & + & + & - & -\\
O\!\mapsto\!UOU^\dagger: & - & + & - & + & + & + & - & -\\
O\!\mapsto\!UO^TU^\dagger: & - & - & - & + & + & + & + & +
\end{array}
\label{eq:S:table}
\end{equation}
The last row gives the time-reversal pattern $S_i\to-S_i$, $T_{jk}\to T_{jk}$: on the $B$ factor of any product state $\mathcal U_B\circ T_B$ reproduces $\Theta\rho_B\Theta^{-1}$ exactly, which is why we call it partial time reversal, the standard reading of partial transposition~\cite{HorodeckiRMP2009}. It is linear but, like $T_B$ itself, positive rather than completely positive, so it is not a physical channel --- precisely what enables entanglement detection. Since the $\Lambda$-side operators are untouched, the composition flips exactly $P_i$ and $C_{ij}$ --- the inaccessible sector of Sec.~\ref{sec:sm_4} --- and fixes all 23 accessible parameters. Averaging over $\mathbb Z_2=\{\mathds{1},\mathcal U_B\circ T_B\}$ therefore yields Eq.~\eqref{eq:meas} of the main text,
\[
\rho_{\rm meas}=\tfrac12\big[\rho_{\rm true}+\mathcal U_B(\rho_{\rm true}^{T_B})\big],
\]
as follows directly by applying the flip pattern of Eq.~\eqref{eq:S:table} to all 35 parameters.

\medskip\noindent\emph{Proof of Lemma 1.}\ \ Every parameter surviving in $\rho_{\rm meas}$ is either invariant under both $T_B$ and $\mathcal U_B$ (the 15 routinely measured ones) or flipped by both ($I_{10},I_{1m},L_{i,\alpha}$), as seen from Eq.~\eqref{eq:S:table}. In either case $T_B$ and $\mathcal U_B$ act identically on every operator present in $\rho_{\rm meas}$, so $\rho_{\rm meas}^{T_B}=\mathcal U_B(\rho_{\rm meas})$. Since $\mathcal U_B$ is a unitary conjugation, they share a spectrum.

Theorems~1 and~2 are proved in the main text. One consequence for the two sections below is worth stating: when an entangled state escapes detection, this is not a reconstruction artifact: some separable state reproduces every measured number, so no method can certify entanglement from those data. 

\medskip\noindent\emph{Verification of Proposition 1 (pure entangled states).}\ \ Any pure state of a $2\times3$ system has Schmidt rank $\le\min(2,3)=2$ and is entangled iff the rank is exactly 2: $|\psi\rangle=\sqrt{\lambda_1}|1\rangle_A|1\rangle_B+\sqrt{\lambda_2}|2\rangle_A|2\rangle_B$. For $|1\rangle_B=|{+1}\rangle$, $|2\rangle_B=|0\rangle$ --- the Clebsch--Gordan-motivated basis of Sec.~\ref{sec:sm_2} --- direct computation gives $\lambda_{\min}(\rho_{\rm meas}^{T_B})=-\tfrac12\sqrt{\lambda_1\lambda_2}$ for all $\lambda_1,\lambda_2>0$, exactly half of $\lambda_{\min}(\rho_{\rm true}^{T_B})=-\sqrt{\lambda_1\lambda_2}$. The criterion is simple: since $U$ is real and symmetric with $U^2=\mathds{1}$, one has $\langle u|U|v\rangle=\langle v|U|u\rangle^{*}$, and the factor is exactly $\tfrac12$ whenever $\alpha\equiv\langle 2|_BU|1\rangle_B$ vanishes, i.e.\ whenever the two Schmidt vectors are not connected by the $\pi$ rotation $U$. This includes the Clebsch--Gordan basis and the full rotated real family $|1\rangle_B=\cos s|{+1}\rangle+\sin s|{-1}\rangle$, $|2\rangle_B=|0\rangle$, independently of $s$.

\medskip\noindent\emph{Remark: numerical evidence on pure-state detection.}\ \ The factor $\tfrac12$ is not universal: over randomly generated Schmidt bases the ratio $\lambda_{\min}(\rho_{\rm meas}^{T_B})/(-\sqrt{\lambda_1\lambda_2})$ varies, dipping to $\approx0.11$ near the separable boundary. $\lambda_{\min}$ was nevertheless strictly negative in all $3\times10^4$ random pure entangled states tested (random complex Schmidt vectors on both sides, $\lambda_1$ uniform on $(0.001,0.999)$), the closest approach to zero being $-4.7\times10^{-3}$ at $\lambda_1=0.999$. A global numerical maximization of $\lambda_{\min}(\rho_{\rm meas})$ over pure entangled states, with the Schmidt weight and both Schmidt bases free, likewise found no non-negative value. We therefore \emph{conjecture}, without proof, that $\rho_{\rm meas}$ detects every pure entangled state of this system. The natural single witness --- the negative eigenvector of $\rho_{\rm true}^{T_B}$ mapped through $\mathcal U_B$ --- does not suffice: its expectation value depends on $\alpha$, and it ceases to certify for Schmidt bases with $|\alpha|$ near unity, where detection is instead supplied by a different eigenvector. By Theorem~2, the conjecture is equivalent to stating that no pure entangled state of this system shares its 23 accessible parameters with any separable state. Note that the ratio above tends to zero as $\lambda_1\to0$, as it must: the state becomes separable in that limit.

\medskip\noindent\emph{Verification of Proposition 2 (mixed-state blind windows).}\ \ Mixed states exhibit genuine blind windows. We consider two families: dilution by noise and, more strikingly, a mixture of two \emph{entangled} states.

\emph{Dilution by noise.} Dilute the maximally entangled state $|e_0^+\rangle$ of Eq.~\eqref{eq:S:bell} with isotropic noise, $\rho_{\rm true}(p)=p|e_0^+\rangle\langle e_0^+|+\tfrac{1-p}6\mathds{1}_6$, $p\in[0,1]$. Direct symbolic evaluation gives
\[
{\rm spec}\big(\rho_{\rm true}(p)^{T_B}\big)=\Big\{\tfrac16+\tfrac p3\,({\times}3),\ \tfrac16-\tfrac{2p}3,\ \tfrac16-\tfrac p6\,({\times}2)\Big\},
\]
so $\rho_{\rm true}(p)$ is entangled iff $p>1/4$ (only $\tfrac16-\tfrac{2p}3$ can go negative on $[0,1]$). This family populates only the four inaccessible parameters $P_z,C_{xx},C_{yy},C_{zz}$, and
\begin{multline*}
{\rm spec}\big(\rho_{\rm meas}(p)^{T_B}\big)=\Big\{\tfrac16{+}\tfrac p{12}({\times}2),\ \tfrac16{-}\tfrac{5p}{12},\ \tfrac16{-}\tfrac p6,\\
\tfrac16{+}\tfrac{5p}{24}{\pm}\tfrac{\sqrt5p}8\Big\},
\end{multline*}
whose smallest member $\tfrac16-\tfrac{5p}{12}$ turns negative only for $p>2/5$ (the remaining five entries are non-negative on $[0,1]$ by inspection, including the lower $\pm$ branch, which equals $\tfrac16-\tfrac{3\sqrt5-5}{24}p>\tfrac{9-3\sqrt5}{24}>0$ throughout). Thus for $p\in(\tfrac14,\tfrac25)$ the true state is entangled while $\rho_{\rm meas}(p)\succeq0$; by Lemma~1 the reconstruction is then PPT, hence separable, and reproduces the data by construction, so Theorem~2 establishes the claim. For example, at $p=0.3$: $\lambda_{\min}(\rho_{\rm true}^{T_B})=-\tfrac1{30}<0$ (entangled) while $\lambda_{\min}(\sigma^{T_B})=+\tfrac1{24}>0$ for $\sigma\equiv\rho_{\rm meas}(0.3)$ (separable). The two states agree on all 23 accessible parameters and differ only in $P_z=\tfrac9{40}$, $C_{xx}=\tfrac{9\sqrt2}{40}$, $C_{yy}=-\tfrac{9\sqrt2}{40}$, $C_{zz}=\tfrac9{40}$, all zero in $\sigma$.

\emph{Mixtures of two entangled states.} Take $\rho_{\rm true}(p)=p|e_0^+\rangle\langle e_0^+|+(1-p)|e_1^+\rangle\langle e_1^+|$, a mixture of two maximally entangled states with no separable component. Its partial transpose carries the eigenvalue $\tfrac p4-\tfrac14\sqrt{5p^2-8p+4}$, negative for every $p\in[0,1)$ since $5p^2-8p+4-p^2=4(1-p)^2>0$, and $-\tfrac12$ at $p=1$: so the state is entangled throughout. The reconstruction has smallest eigenvalue
\[
\min\Big\{\tfrac{3p}8-\tfrac18\sqrt{17p^2{-}16p{+}4},\ \tfrac38{-}\tfrac{3p}8{-}\tfrac18\sqrt{17p^2{-}18p{+}5}\Big\},
\]
the first branch negative only for $p<1-\tfrac1{\sqrt2}$ and the second only for $p>\tfrac1{\sqrt2}$ (both radicands are positive throughout, discriminants $-16<0$). Thus entanglement is uncertifiable for $p\in[1-\tfrac1{\sqrt2},\tfrac1{\sqrt2}]$, a window of width $\sqrt2-1\approx0.414$; at the midpoint $p=\tfrac12$, $\lambda_{\min}(\rho_{\rm true}^{T_B})=\tfrac{1-\sqrt5}8\approx-0.155$ while $\lambda_{\min}(\rho_{\rm meas}^{T_B})=+\tfrac18$. The two endpoints, pure Bell states, are correctly detected with $\lambda_{\min}=-\tfrac14$ each (Proposition~1); only their mixture loses the signal in the accessible sector. A scan of 300 random mixtures of two random entangled pure states produced no blind window, indicating that this is a structured feature of this pair rather than generic behaviour, and showing that no bound of the form ``entangled enough implies certifiable'' is available.

\bibliographystyle{apsrev4-2}
\bibliography{QQQT_draft}

\end{document}